# Which K-Space Sampling Schemes is good for Motion Artifact Detection in Magnetic Resonance Imaging?

Mohammad Reza Mohebbian[1], Ekta Walia[2], Khan A. Wahid[1]

*Abstract*— Motion artifacts are a common occurrence in the Magnetic Resonance Imaging (MRI) exam. Motion during acquisition has a profound impact on workflow efficiency, often requiring a repeat of sequences. Furthermore, motion artifacts may escape notice by technologists, only to be revealed at the time of reading by the radiologists, affecting their diagnostic quality. Designing a computer-aided tool for automatic motion detection and elimination can improve the diagnosis, however, it needs a deep understanding of motion characteristics. Motion artifacts in MRI have a complex nature and it is directly related to the k-space sampling scheme. In this study we investigate the effect of three conventional k-space samplers, including Cartesian, Uniform Spiral and Radial on motion induced image distortion. In this regard, various synthetic motions with different trajectories of displacement and rotation are applied to T1 and T2-weighted MRI images, and a convolutional neural network is trained to show the difficulty of motion classification. The results show that the spiral k-space sampling method get less effect of motion artifact in image space as compared to radial k-space sampled images, and radial k-space sampled images are more robust than Cartesian ones. Cartesian samplers, on the other hand, are the best in terms of deep learning motion detection because they can better reflect motion.

*Clinical Relevance*—This study helps to understand the impact of different k-space samplers on motion artifact. The outcome of this study can be used for designing an artificial intelligence system for motion detection or correction, taking the k-space scheme into consideration and also may lead to improvement in MRI designs.

## I. Introduction

Magnetic Resonance Imaging (MRI) is frequently acquired to plan for surgery [1], investigate the effect of drug treatment [2] and study diseases such as cancer [3], Alzheimer [4], and Huntington's disease [5]. MRI allows precise anatomical, physiological and functional assessment non-invasively [6]. The variety of applications in MRI has promoted awareness of image quality and image acquisition techniques.

MRI has a complex nature and acquisition time is potentially long in which the image quality can be affected by motion artifact easily [7]. Motion can reduce image quality due to blurring, shading, and structured artifacts that reduce reliability. As an instance, head motion in MRI images can change cortical thickness and cause the unreliability of images in clinical applications. Motion artifact in an image decreases the measured values of gray matter. Even a minimal motion can lead to a decrease in the gray matter values by over 4% [8].

There are various techniques to detect or decrease the effect of motion on MRI images. As an instance, mechanical foam padding around the subject's head can reduce mild movement [9]. A synchronized sensor for tracking the subject's movement during image acquisition is another example of motion detection which can eliminate the misinterpretation in image assessments [10]. However, the calibration time for these sensors is long, and a change in MRI MRI device design might be needed for embedding on traditional hardware and software. Experts also can inspect image quality after acquisition; however, the availability of experts and time is another issue. This problem can be eliminated by machine learning and image processing approaches [11]. Computer-aided diagnosis with image processing and machine learning technique can keep the examination time and cost low. Furthermore, if an algorithm can detect motion during acquisition time it can trigger a partial re-scan instead of scanning the whole volume.

The initial step for training a machine learning model to classify the motion artifacts is to understand how motion appears on images. The way motion appears on MRI images is highly dependent on device configuration and image acquisition techniques. The images are registered in the k-space domain with the k-space sampler and the effect of k-space sampler on motion artifacts needs more investigation.

The focus of this paper is to investigate the impact of k-space sampler trajectories like Radial, Cartesian and Spiral sampling on the appearance of motion in the image domain. A toolbox is designed for creating synthetic motion on MRI images with real-world parameters, such as Time Repetition (TR), the number of excitation (NEX), motion displacement and rotation, k-space sampler speed etc. One of the purposes of this work is to investigate whether motion classification on different k-space sampled images exhibits similar performance. Also, this research helps to show which k-space sampling technique is less sensitive to motion. Furthermore, the designed toolbox opens a new area for further analysis of the various k-space schemes.

The rest of the paper is organized as follows: in the next section, information about the images and formulation of methods used in this study is presented. Section 3 provides the

1 Department of Electrical and Computer Engineering, University of Saskatchewan S7N 5A9, Saskatoon, Saskatchewan, Canada
2 Advanced Innovation, Enterprise Operational Informatics, Philips HealthCare, 281 Hillmount Road, L6C2S3, Markham, Ontario, Canada.

Corresponding Author: MohammadReza Mohebbian, Email:mom158@usask.ca, Tel: (+1)6393181602

## II. Materials and Methods

### A. Generating synthetic motion based on the k-space trajectory

In this paper, retrospective motion simulation is used to generate synthetic motion artifacts on MR images. One of the assumptions is that the subject volume behaves as a rigid body, hence, we can create arbitrary motion trajectories in six degrees of freedom for displacement and rotation using piecewise functions.

If the subject volume is rigidly moving, and movement is piecewise constant, it implies that the subject does not move very fast during a limited time. The subject volume during each motion, displacement and rotation is based on Equation 1.

$$I_t = A_t I_{t-1} \quad (1)$$

Where, $A_t$ is transformation-rotation tensor in time $t$ respect to the time $t-1$, and $I_t$ is the subject volume with size $N \times M \times P$ in pixel space at time $t$. Since the scanner time is limited, only part of the k-space is being filled at time $t$. The scan time in MRI is typically dependent on TR, NEX and phase step. The TR measures the time from one excitation pulse to the next in milliseconds. The phase step in the image matrix will determine the number of lines in k-space which should be filled. The more lines exist in k-space, the more echoes will be collected. The scan time is long and as an instance, for a spin-echo sequence with TR 400 ms, matrix size $208 \times 256$ and NEX 2, the scan time is 166 seconds. Each phase encoding step requires 1 TR interval. Therefore, for this example 208 TR interval should pass for one excitation.

For generating the motion artifact on slices, a movement trajectory, including six degrees of freedom is defined randomly from time 0 to the end of the acquisition time. The root means square (RMS) of displacement and rotation of the motion trajectory can reflect the motion severity, which is set to $1 \pm 0.4$ mm displacement and $0.6 \pm 0.4$ rotation. Since the trajectory for every six degrees of freedom is defined randomly, the RMS of displacement and rotation are reported based on average and standard deviation. Random trajectories generated in the experiments are smoothed using a sgolay filter to be more like a patient movement in hospital setting. In this regard, high variations are smoothed and severe sudden movement is controlled. A scanner is defined according to the typical scanner parameters. Then, the $I_t$ volume is calculated for every time $t$ and the corresponding region of interest (ROI) of k-space for a particular slice is sampled by the scanner. Selecting the ROI of a slice in k-space is done according to the k-space sampler timestamp. By assuming that image has size 256×256×89, one random trajectory can be defined as shown in Figure 1.

### B. Dataset

In this paper, we use the IXI brain development database [12] which is an open-access MRI image dataset. The IXI database has T1, T2, and PD-weighted brain images and is collected at three different hospitals in London with 1.5 and 3 Tesla systems using Philips and GE devices. The T1-weighted slices have $208 \times 208$ matrix size, with 9.6 ms TR, 4.6 TE and flip angle 8 degrees. While the T2-weighted images have size $192 \times 187$ with TR 5725 ms and TE 100 ms. The flip angle is 90 degrees. For simplicity, all slices are resized to 256×256.

Since characteristics of T1 and T2-weighted images are different, the dataset is separated into two parts consisting of T1-weighted and T2-weighted images. The specification of the final slices in each subset is provided in Table 1.

TABLE 1. SUMMARY OF THE DATASETS.

| Type of images | Total number of images |
|---|---|
| (a) T1-weighted | 44520 |
| (b) T2-weighted | 41760 |

### C. Deep learning for motion classification

Both the datasets were independently used to train two convolutional neural network (CNN) models from scratch. Each CNN model performed preprocessing of the images, including normalization and resizing to $256 \times 256$, as the first step. The CNN architecture consisted of fifteen layers including three convolution layers. The first layer was the input layer configured to $256 \times 256 \times 1$. A convolutional layer (Conv1) consisting of 8 filters of 3x3 sizes was added after the input layer. During convolution operation stride 1 was used so that no feature was missed. The inputs of each layer were normalized in the next layer i.e. the batch normalization layer such that they have a mean output activation of zero and a standard deviation of one. The rectified linear unit was used as an activation layer. As the pooling layer, max-pooling was used with the size of 2x2 with stride 2 which was followed by the second convolutional layer (Conv2) consisting of 16 kernels of size 3x3. A similar configuration for batch normalization, activation layer, and the max-pooling layer was implemented. The last convolutional layer (Conv3) consisted of 64 filters of 3x3 size. After that a fully connected layer was directly added which was configured to two nodes for binary classification. The final classification layer following a soft-max layer was used to compute the cross-entropy loss for classification.

## III. Results

The motion classification performance on Cartesian, Radial, and Uniform spiral k-space sampled images is provided in Table 2. All analyses are performed 5 times in the image domain and results are provided for test-set which is 30 percent of the whole data. In each run, random samples are divided to train and test and CNN models (one for T1-weighted and another for T2-weighted) are applied to them.

The motion classification on Spiral k-space has less sensitivity than two other k-space sampling methods, which means that Spiral k-space sampling is more robust against the motion and does not reflect the motion in a way that can be detected easily. On the other hand, the Cartesian k-space sampling is very sensitive to motion and it will ease the motion classification or reconstruction applications.

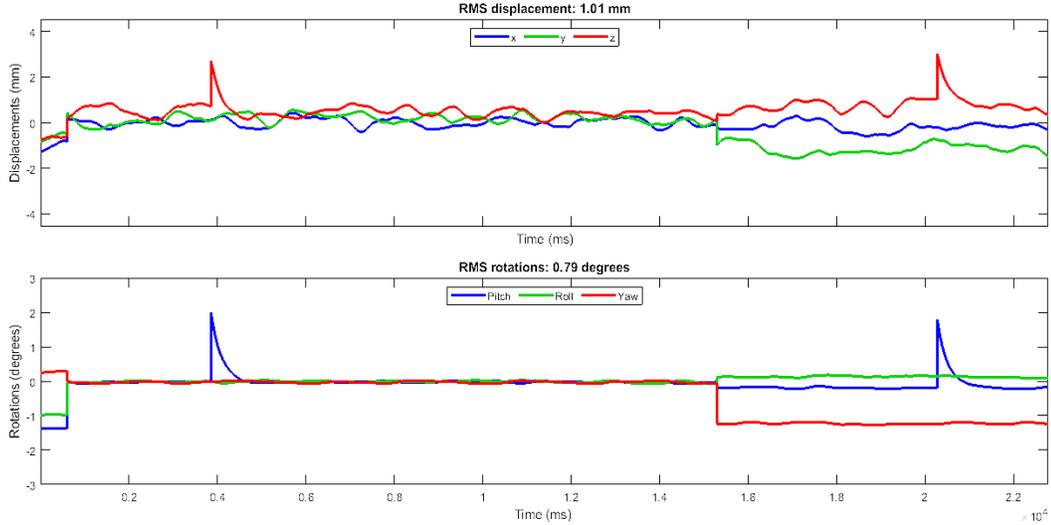
Figure 1. Random trajectory created for the whole acquisition time.

Next, we discuss how the time of motion occurrence is an important factor as well. The k-space is the Fourier transform of MR image and values of k-space corresponding to the spatial frequency of the MR image. Mathematically, the higher frequency components appear in the boundary of k-space which includes detail of the image and low-frequency components which contains the approximation of the image is in the center. If the motion happens in a time that k-space sampler is near the center of k-space, more artifacts are visible on the image and the artifact has low frequency. On the other hand, the motion artifact occurs on the boundary of the image has a higher frequency and it may not appear in the image. Therefore, the time when motion occurs is important. A bulk motion occurring at a certain time may not have an image distortion effect that a little motion occurring at another time can have. Hence, the motion distortion effect pertains to both time and k-space sampling scheme.

Figure 2 shows the effect of motion in different k-spaces. It also shows the effect of time of motion occurrence on motion artifact. The k-space is like two-dimensional fourier that represents spatial frequencies. Depicted k-space is like shifted fourier transform of image, wherein the center has low-frequency information and edges have the highest frequency. It can be observed that the distortion in low frequency is smaller, however, it has a severe effect on the image.

## IV. DISCUSSION

Motion artifacts are mostly a problem when they obscure some anatomy of interest. Thus, one solution is to switch the direction of phase encoding. In non-Cartesian k-space sampling, instead of sampling k-space one line at a time in a rectangle, the scanner will sample lines or small rectangles of different angles across k-space. Because phase encoding occurs in different directions for each line, that tends to minimize the motion artifacts in the image. According to Table 2, the sensitivity on Cartesian k-space sampled is better than radial k-space sampled images and radial has better sensitivty than uniform spiral. The more sensitivty is achieved by Cartesian sampling reveals that this method reflects motion explicitly. On the other hand, the radial sampling works on symmetric lines according to the k-space center which reduces the blurring and noisy effect in image with respect to other sampling methods.

The way that motion distortion is related to the k-space sampler scheme can be elaborated by gradient direction. Gradients along the x-axis or frequency-encode direction make blurring and shearing effect, which is proportional to the scan time. On the other hand, gradients along the y-axis or phase-encoded direction can change the effective time echo. This direction has a lower bandwidth than other directions and may displace off-resonance spins like fat protons. In addition, gradients along the z-axis or slice-select direction produce void signals and phase the signal. In this regard, the Cartesian k-space sampling has more blurring and shearing artifact and is the easiest way to implement which been the standard method for most MR imaging sequences and echo-planar imaging for a long time. Radial and spiral readouts offer low intrinsic sensitivity to motion and have lower time echo which has different artifacts, including ring-shaped blurring and curvilinear bands because they are not aligned with the x and y-axis.

TABLE 2. MOTION DETECTION PERFORMANCE USING CNN. THE AVERAGE AND STANDARD DEVIATION FOR 5 RUNS ARE REPORTED.

| Models (trained/tested) | k-space sampling | Accuracy | Recall (Sensitivity) | Specificity | Precision | F1-Score |
|---|---|---|---|---|---|---|
| T1-weighted images | Cartesian | 100 ± 0.0 | 100 ± 0.0 | 100 ± 0.0 | 100 ± 0.0 | 100 ± 0.0 |
| | Spiral | 98.9 ± 0.4 | 98.7 ± 0.5 | 99.1 ± 0.3 | 99.0 ± 0.4 | 98.9 ± 0.4 |
| | Radial | 99.3 ± 0.5 | 99.1 ± 0.6 | 99.5 ± 0.5 | 99.5 ± 0.6 | 99.3 ± 0.6 |
| T2-weighted images | Cartesian | 100 ± 0.0 | 100 ± 0.0 | 100 ± 0.0 | 100 ± 0.0 | 100 ± 0.0 |
| | Spiral | 98.5 ± 0.8 | 98.3 ± 0.9 | 98.6 ± 0.7 | 98.5 ± 0.8 | 98.4 ± 0.8 |
| | Radial | 98.9 ± 1.0 | 98.9 ± 1.2 | 99.0 ± 0.3 | 98.7 ± 0.9 | 98.9 ± 0.9 |

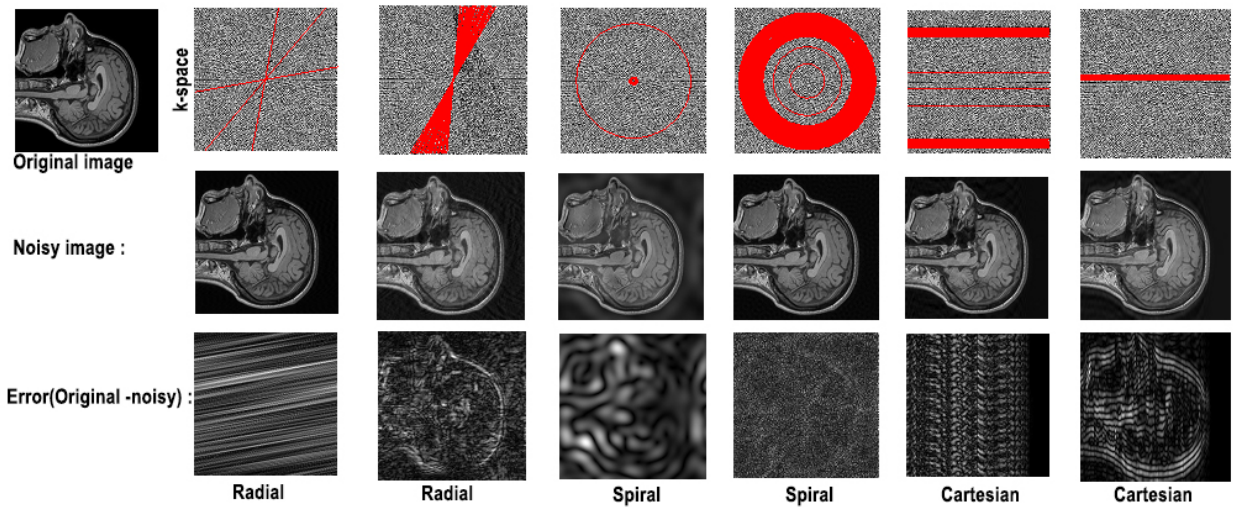

Figure 2. Motion artifact in various k-space sampling schemes. The red-line in k-space denotes the line of the k-space when motion happened and k-space is distorted. The absolute error is the absolute value of difference between the original and noisy images. This error is normalized to a 8-bit grayscale to be visible.

## V. Conclusion

Non-Cartesian k-space sampling, including spiral and radial sampling, reflects less motion on images and is more robust against motion. This concept is proved by this research in which the motion classification algorithm has less sensitivity when they are trained with images registered by the non-Cartesian k-space sampler.

A convolutional neural network is trained to show the difficulty in motion classification of non-Cartesian sampled images. The spiral k-space sampling method has least effect of motion in image space. Radial samplers are observed to be more robust than Cartesian ones. Also it is shown that time of motion occurence during k-space sampling is also a paramount factor in determining the extent of image distortion.

## Acknowledgment

We thank the National Sciences and Engineering Research Council of Canada (NSERC) for supporting this work through the Engage grant